\documentclass[11pt, a4paper]{article}

\usepackage{geometry}
\usepackage{amsmath}
\usepackage{amssymb}
\usepackage{graphicx}
\usepackage{bm}
\usepackage{color}
\usepackage[colorlinks=true, linkcolor=blue, filecolor=magenta, urlcolor=blue, citecolor=blue]{hyperref}
\usepackage{authblk} 

\title{Geometric Matching of Local Static Regions in Cosmological Spacetimes with an Evolving Lapse}

\author{Seokcheon Lee\thanks{E-mail: skylee@skku.edu}}
\affil{\textit{Department of Physics, Institute of Basic Science, Sungkyunkwan University, Suwon 16419, Korea}}

\date{\today}

\begin{document}

\maketitle

\begin{abstract}
The generalized cosmological time (GCT) framework introduces a modified lapse function, $N(t)\propto a^{b/4}$, as a geometric extension of the standard FLRW description. Like other departures from $\Lambda$CDM, such constructions must remain compatible with the observed stability of local gravitational and laboratory physics. In scalar--tensor theories, this compatibility is usually achieved through dynamical screening mechanisms that suppress additional degrees of freedom in dense environments.  In this work, we examine whether a locally static spacetime region can be consistently embedded within a cosmological background described by a time--dependent lapse.  By embedding a static Schwarzschild interior into an expanding GCT--FLRW exterior, the Israel junction conditions are used to determine the class of background expansions that admit such a matching in the absence of a thin shell. The continuity of the extrinsic curvature yields a Friedmann--type relation that coincides with the GCT background equations of motion.  This relation should be interpreted not as a new dynamical equation, but as a geometric consistency condition (GCC) associated with the matching of the two spacetime regions. In this sense, the junction does not introduce new dynamics, but provides a GCC under which a region with fixed local clocks can be embedded in a cosmological spacetime with an evolving lapse.  Therefore, the analysis clarifies how locally static gravitational systems can remain compatible with a cosmological time normalization that differs from that of local proper time while preserving the standard description of local physics.
\end{abstract}

\section{Introduction}
\label{sec:intro}

The standard $\Lambda$CDM model successfully accounts for a wide range of cosmological observations, including cosmic microwave background (CMB) anisotropies \cite{Planck:2018nkj,Planck:2018vyg}, large--scale structure (LSS) measurements \cite{SDSS:2005xqv,SDSS:2008tqn}, and baryon acoustic oscillation surveys \cite{BOSS:2014hwf,BOSS:2016wmc,eBOSS:2020yzd,eBOSS:2020fvk}. With the increasing precision of modern data, however, statistically significant discrepancies have emerged between early-- and late--Universe measurements. The most prominent is the Hubble tension, in which the value of the Hubble constant inferred from the CMB differs from that obtained from local distance--ladder observations \cite{DiValentino:2021izs,Riess:2021jrx}. Related anomalies, such as the $S_8$ tension in the amplitude of matter clustering, have further motivated the exploration of extensions to the standard cosmological framework \cite{Heymans:2020gsg,DES:2021wwk}.

A broad range of theoretical approaches has been proposed to address these tensions, including early dark energy models \cite{Poulin:2018cxd,Smith:2019ihp,Hill:2020osr} and modifications of gravity \cite{DiValentino:2016hlg,Ivanov:2019pdj,Yang:2021flj,Perivolaropoulos:2021jda,Kamionkowski:2022pkx}.  Among these, the generalized cosmological time (GCT) framework, whose minimally extended varying speed of light (meVSL) realization was developed in \cite{Lee:2020zts,Lee:2023bjz,Lee:2025osx,Lee:2023ucu,Lee:2024kxa}, introduces a geometric modification of the Friedmann--Lema\^{i}tre--Robertson--Walker (FLRW) metric through an evolving lapse function $N(t)$. In this construction the lapse is parameterized as $N(t)=a^{b/4}$, which introduces modified cosmological time--dilation effects while preserving the standard treatment of early--Universe physics such as Big Bang Nucleosynthesis and the CMB acoustic scale \cite{Lee:2022heb,Lee:2025vha}. Within this framework the normalization of cosmological time is allowed to differ from that of locally measured proper time, providing a geometric setting in which cosmological time dilation (CTD) can deviate from the standard FLRW scaling while local physics remains unchanged \cite{Lee:2026kyz}.

Any extension that modifies the relation between cosmological time and physical observables must remain consistent with the stringent bounds imposed by local experiments and astrophysical tests. Solar--system dynamics, binary pulsars, and laboratory measurements tightly constrain possible variations of fundamental constants, including limits on $\dot G/G$ \cite{Uzan:2010pm,Will:2014kxa}. These constraints raise the question of how a cosmological modification of time normalization can remain compatible with the observed stability of local gravitational and laboratory physics.

In scalar--tensor and related modified--gravity theories, this compatibility is usually achieved through dynamical screening mechanisms, such as the chameleon or Vainshtein effects, which suppress additional degrees of freedom in high--density environments \cite{Khoury:2003aq,Khoury:2003rn,Vainshtein:1972sx}. The role of time slicing and lapse functions in general relativity has been discussed extensively in the literature~\cite{Wald:1984rg,Gourgoulhon:2012ffd,Ellis:1971pg}. The present work instead examines a geometric aspect of this problem by studying whether a locally static spacetime region can be consistently embedded within a cosmological background described by a different time normalization.

Using the $3+1$ formulation of the GCT model \cite{Lee:2024zcu} together with the Israel junction conditions (IJCs)~\cite{Israel:1966rt}, this work considers a composite spacetime in which a static Schwarzschild interior is embedded within an expanding GCT--FLRW exterior. The requirement that the extrinsic curvature be continuous across the matching hypersurface, in the absence of a thin shell, leads to a Friedmann--type relation that coincides with the GCT background equations. This relation should be interpreted not as a new independent dynamical equation, but as a geometric consistency condition (GCC) associated with the matching of the two spacetime regions. Thus, the junction identifies the class of background evolutions that are compatible with the coexistence of a locally static region and a cosmological spacetime with an evolving lapse.

Within this construction, the interior region admits a timelike Killing vector normalized by a unit lapse, so that local notions of time and energy coincide with those of standard general relativity. The exterior region, by contrast, is described by a cosmological time coordinate whose normalization evolves with the scale factor. The matching conditions ensure that these two descriptions can be embedded in a single spacetime without introducing surface stresses. This separation between a local static gauge and a global cosmological gauge provides a geometric framework for analyzing CTD effects while maintaining the standard description of local physics.

The purpose of the present paper is to establish this geometric consistency at the level of spacetime matching. Issues related to the dynamical formation of such regions and to detailed phenomenological applications are deferred to companion studies.

The paper is organized as follows. Section~\ref{sec:formalism} reviews the $3+1$ formulation of the GCT model and the role of the modified lapse function. Section~\ref{sec:junction} presents the metric junction conditions between the static interior and the evolving exterior. In Section~\ref{sec:derivation}, the equations of motion for the boundary shell are derived and shown to imply the GCT-modified Friedmann equation. Sections~\ref{sec:discussion} and~\ref{sec:conclusion} discuss the implications for virialized systems, black-hole spacetimes, and the consistency of physical laws across different time gauges.

\section{GCT Formalism and 3+1 Decomposition}
\label{sec:formalism}

The GCT framework explores the possibility that the
speed of light $c$, when expressed with respect to the global cosmological time
coordinate $t$, may evolve with the scale factor $a(t)$~\cite{Lee:2024zcu}.
This departure from the standard FLRW description is represented by the line element
\begin{equation}
ds^2 = -c(t)^2 dt^2 + a(t)^2 \gamma_{ij} dx^i dx^j \,,
\label{eq:GCT_metric}
\end{equation}
where $\gamma_{ij}$ denotes the metric on spatial hypersurfaces of constant curvature.
To maintain consistency with locally measured constants and standard units, the temporal
coordinate is taken to be $x^0=c_0 t$, with $c_0$ the present-day locally measured speed
of light.

To analyze the causal structure and the evolution of spatial hypersurfaces, the
Arnowitt--Deser--Misner (ADM) $3+1$ decomposition is employed~\cite{Arnowitt:1962hi}.
In this formalism, the spacetime metric is written as
\begin{equation}
ds^2=-N^2(dx^0)^2+h_{ij}(dx^i+N^i dx^0)(dx^j+N^j dx^0)\,,
\end{equation}
where $N$ is the lapse function, $N^i$ the shift vector, and $h_{ij}$ the induced spatial
metric.
Comparing this form with Eq.~(\ref{eq:GCT_metric}) and using $dx^0=c_0 dt$, one finds that
for comoving observers $N^i=0$, while the lapse is
\begin{equation}
N(t)\equiv \frac{c(t)}{c_0}\,.
\label{eq:lapse_def}
\end{equation}

Within the meVSL realization of GCT, the lapse is assumed to follow a simple power-law dependence on the scale factor~\cite{Lee:2020zts},
\begin{equation}
c(t)=c_0\,a^{b/4}(t)
\qquad\Longrightarrow\qquad
N(t)=a^{b/4}(t)\,,
\label{eq:GCT_ansatz}
\end{equation}
where the dimensionless parameter $b$ quantifies deviations from the standard FLRW case,
recovered for $b=0$.

As emphasized in standard treatments (e.g., Ryder~\cite{Ryder:2009}), the Cosmological
Principle and Weyl’s postulate restrict the FLRW metric component $g_{00}$ to be an
arbitrary function of time, $N(t)$, but do not fix it to unity.
In conventional cosmology, time reparameterization invariance is typically used to set
$N(t)=1$ globally, identifying the coordinate time with the proper time of comoving
observers.
This choice, however, implicitly assumes that the normalization of cosmological time
coincides everywhere with that of local physical clocks.

In the GCT framework, the lapse function is treated as a geometric degree of freedom
encoding the relative normalization between cosmological time and local proper time.
In this interpretation, locally virialized systems may be described by a
static time gauge, while the cosmological background is characterized by an evolving
lapse. Therefore, retaining $N(t)$ explicitly in the metric preserves physical information about
the mismatch between local and global time normalizations, which manifests in
cosmological observables such as time dilation and luminosity distances
\cite{Lee:2020zts,Lee:2025osx,Lee:2023ucu,Lee:2024kxa}.

The consistency of the Einstein equations and the Bianchi identity in the meVSL/GCT
framework has been established directly from the Einstein--Hilbert action
\cite{Lee:2020zts,Lee:2025osx}.
In particular, the coupled evolution of $c(t)$ and $G(t)$ is constrained such that the
Einstein gravitational coupling $\kappa\equiv 8\pi G/c^4$ remains constant, ensuring the
standard conservation law $\nabla_\mu T^{\mu\nu}=0$.

Unlike scalar--tensor theories, where lapse modifications are typically sourced by local
scalar fields coupled to matter, the lapse in the GCT framework depends only on the global
scale factor. Thus, the modification is global and geometric rather than locally sourced.

A central question addressed in this work is whether a region characterized by a static
lapse ($N=1$) can be consistently embedded within a cosmological background in which
$N(t)=a^{b/4}$.
This issue cannot be resolved within a single coordinate patch, where the lapse may
always be absorbed by a time reparameterization.
Instead, the present construction considers a composite spacetime composed of regions
with inequivalent physical time normalizations, joined across a timelike hypersurface.
Thus, the purpose of the junction analysis presented below is to
identify the geometric conditions under which such regions can be consistently matched.
As shown below, the IJCs then constrain the relative normalization
of time between these regions, allowing a locally static time gauge to be embedded
consistently within an evolving cosmological background without introducing additional
propagating degrees of freedom.

\subsection{Global time atlas obstruction}

The distinction between local and cosmological time normalizations becomes manifest when
the spacetime is described by multiple coordinate charts joined across a hypersurface.
Consider a spacetime manifold $\mathcal M$ covered by two charts,
$\mathcal U_-$ (interior) and $\mathcal U_+$ (exterior), matched across a timelike
hypersurface $\Sigma$.
In the interior region, the time coordinate $t_-$ is chosen to coincide with the proper
time of static observers, so that $g_{t_-t_-}=-1$.
In the exterior region, the time coordinate $t_+$ is identified with the cosmological
time parameter governing the Hubble flow, with $g_{t_+t_+}=-N^2(t_+)$.

On the hypersurface $\Sigma$, the proper time $\tau$ is geometrically defined and must be
unique,
\begin{equation}
c_0^2 d\tau^2 = - g_{\mu\nu} dx^\mu dx^\nu \Big|_{\Sigma}.
\end{equation}
Therefore,  consistency requires that the transition functions between the two charts
satisfy
\begin{equation}
dt_- = N(t_+) \, dt_+ \qquad \text{on } \Sigma.
\label{atlas}
\end{equation}

If $N(t_+)$ is time dependent, this relation fixes a nontrivial, time-dependent relative
normalization between the interior and exterior clocks.
While each region separately admits a reparameterization that sets its lapse to unity,
no single global time coordinate can simultaneously preserve the physical normalization
of time in both regions once these identifications are imposed.

In this sense, the lapse mismatch reflects the presence of distinct
physical time calibrations associated with the two coordinate patches rather than a
breakdown of coordinate freedom.

Then, the IJCs determine how these calibrations may be consistently
matched, ensuring that a locally static time gauge can coexist with an evolving
cosmological background without introducing additional dynamical degrees of freedom.

\section{Metric Junction Conditions}
\label{sec:junction}

To assess the compatibility of the GCT framework with local gravitational physics, the IJCs are used to match a static interior spacetime to a time-dependent cosmological background. In the present paper we adopt a comoving timelike boundary, $\chi=\chi_b=\mathrm{const}$, so that the physical areal radius of the matching surface is
\begin{equation}
R(\tau)=a\bigl(t_+(\tau)\bigr)\chi_b \, .
\end{equation}
This immediately shows that, under the comoving-boundary ansatz, the boundary worldtube is generically dynamical rather than strictly static. The construction should therefore be interpreted as an idealized zeroth-order geometric matching model, not as a complete model of a fully realistic virialized object with a fixed physical radius.

This point also clarifies the relation to Einstein--Straus or mass-compensation matching. In the standard Einstein--Straus vacuole one likewise has an expanding areal radius,
\begin{equation}
R_\Sigma(t)=a(t)\,r_b \, ,
\end{equation}
together with the mass-compensation condition
\begin{equation}
M=\frac{4\pi}{3}\rho\,R_\Sigma^3 \, .
\end{equation}
Accordingly, the present comoving construction is closer in geometric spirit to an expanding vacuole boundary than to the exact worldtube of a truly static bound system. What is different here is not the existence of spherical matching itself, but the fact that the interior and exterior manifolds are described in inequivalent time normalizations, with a static lapse $N_-=1$ embedded into a cosmological background whose lapse evolves as $N_+(t)=a^{b/4}$.

The purpose of the present section is therefore deliberately limited. We do not claim that the comoving-boundary ansatz already solves the full realistic virialized-boundary problem. Rather, we use it to isolate the geometric consistency conditions that arise when a locally static interior is matched to a cosmological exterior with a distinct temporal normalization. In any region where the metric approaches a static Schwarzschild form, the lapse tends to unity, so that horizon-scale processes, atomic clocks, and laboratory experiments are governed by the usual laws of general relativity and quantum field theory \cite{Lee:2020zts,Lee:2023bjz,Lee:2025osx}. Consequently, local measurements are determined by the interior geometry, while the junction analysis constrains how that local static description can be embedded into a cosmological background with evolving lapse.

\subsection{Interior and Exterior Manifolds}

The interior region, $\mathcal{M}^{-}$, representing a local virialized system, is described by a static and spherically symmetric vacuum solution of Einstein’s equations. This region is modeled by the Schwarzschild metric with constant speed of light $c_0$,
\begin{equation}
ds_{-}^2 = -f(r) c_0^2 dt_{-}^2 + \frac{1}{f(r)}dr^2 + r^2 d\Omega^2 \,,
\label{eq:metric_in}
\end{equation}
where $f(r)=1-\frac{2G_0M}{c_0^2 r}$. In this domain the lapse function is unity ($N_{-}=1$) and the gravitational constant $G_0$ is fixed, corresponding to a locally static environment in which standard general relativity is recovered. This solution therefore provides a natural zeroth-order approximation to the geometry of a gravitationally bound system whose internal dynamics have decoupled from the cosmological expansion.

The exterior region, $\mathcal{M}^{+}$, is taken to be the expanding GCT-FLRW universe. Using the coordinate convention $x^0=c_0 t$, the line element is written as
\begin{equation}
ds_{+}^2 = -c(t)^2 dt_{+}^2 + a(t)^2\left(d\chi^2+\chi^2 d\Omega^2\right)\,,
\label{eq:metric_out}
\end{equation}
where $\chi$ is the comoving radial coordinate and the speed of light e

\subsection{The Hypersurface and First Junction Condition}

The two manifolds are separated by a timelike hypersurface $\Sigma$ defined by $\chi=\chi_b$. On $\Sigma$, the induced metric is written as
\[
ds_{\Sigma}^2=-c_0^2 d\tau^2+R(\tau)^2 d\Omega^2 \,,
\]
where $\tau$ denotes the proper time measured by an observer comoving with the boundary. The first IJC, $[h_{ab}]=0$, requires the induced metric to be continuous across $\Sigma$, which yields the matching of the physical radius,
\begin{equation}
R(\tau)=a(t_{+})\chi_b \,.
\label{eq:radius_match}
\end{equation}
Equation~(\ref{eq:radius_match}) makes explicit that, for $\chi_b=\mathrm{const}$, the physical boundary radius follows the background expansion. In this sense the present ansatz does not describe a strictly static boundary with fixed areal radius. Rather, it should be viewed as an idealized expanding matching surface, analogous at the geometric level to Einstein--Straus-type vacuole constructions, while the realistic non-comoving case $\chi=\chi(\tau)$ is left for future work. 

Equating the proper-time interval from the exterior geometry with that on the hypersurface further gives
\begin{equation}
-c_0^2 d\tau^2=-c(t_{+})^2 dt_{+}^2
\quad\Longrightarrow\quad
\frac{dt_{+}}{d\tau}=\frac{c_0}{c(t_{+})}=a^{-b/4}(t_{+})\,.
\label{eq:time_out}
\end{equation}

This relation provides the kinematic mapping between the local proper time defined on the boundary and the cosmological time coordinate associated with the exterior spacetime.

\begin{figure}[htbp]
\centering
\includegraphics[width=0.95\linewidth]{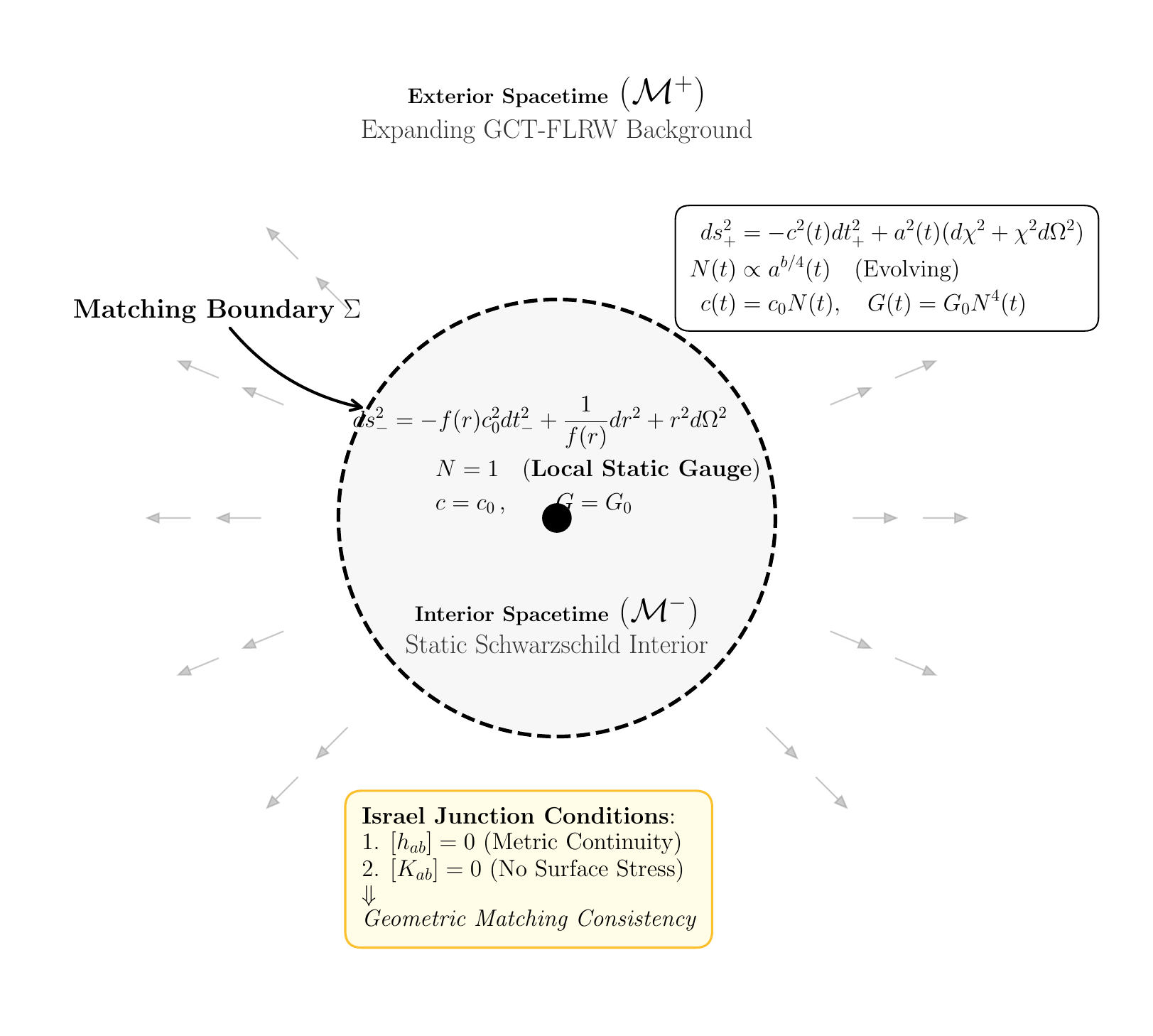}
\caption{Schematic representation of the idealized zeroth-order geometric matching considered in this work. The interior region ($M_-$) is modeled by a static Schwarzschild spacetime with fixed local constants and unit lapse ($N=1$), while the exterior region ($M_+$) is described by an expanding GCT--FLRW background with lapse $N\propto a^{b/4}$. The two regions are joined across a timelike hypersurface $\Sigma$ under the simplifying comoving-boundary ansatz. The IJCs enforce continuity of the induced metric and extrinsic curvature, thereby providing a geometric consistency test for matching distinct temporal normalizations in a single spacetime.}
\label{fig:shielding}
\end{figure}
As illustrated in Fig.~\ref{fig:shielding}, the matching of the two manifolds at $\Sigma$ provides an idealized zeroth-order geometric setup in which a locally static interior and an evolving cosmological exterior can be joined consistently under the comoving-boundary ansatz. The second IJC, $[K_{ab}]=0$, to be discussed in the following section, further constrains the extrinsic curvature across the boundary. The resulting relations should be interpreted as geometric consistency conditions associated with the spacetime matching, rather than as the introduction of a new dynamical screening mechanism.

We stress that the assumption $\chi=\chi_b=\mathrm{const}$ is an idealization. It corresponds to the limit in which the boundary has fully virialized and its comoving position becomes slowly varying. In a more general setting, one should allow $\chi=\chi(\tau)$. Therefore, the present analysis identifies the comoving case as the zeroth--order geometric baseline, from which physically realistic boundaries may deviate~\cite{Lee:PartII}. The matching condition derived below should therefore be interpreted as the leading-order geometric consistency requirement in this limit.

\section{Derivation of the Geometric Consistency Condition}
\label{sec:derivation}

The second IJC requires continuity of the extrinsic curvature tensor across the hypersurface $\Sigma$, $[K_{ab}]=0$. This condition ensures that the matching between the static interior and the expanding exterior is smooth and does not involve a singular surface stress-energy tensor.

\subsection{Extrinsic Curvature Components}

Owing to spherical symmetry, the extrinsic curvature tensor $K_{ab}$ on the hypersurface $\Sigma$ has two independent components: the angular component $K_{\theta\theta}$ (with $K_{\phi\phi}=\sin^2\theta\,K_{\theta\theta}$) and the temporal component $K_{\tau\tau}$. The radial direction is normal to $\Sigma$ and therefore does not contribute to $K_{ab}$. The angular component governs the radial dynamics of the boundary shell, while $K_{\tau\tau}$ is related to surface stresses through the Lanczos equation~\cite{Israel:1966rt}.

Due to the assumption of no surface layer ($S_{ab}=0$), the junction conditions require continuity of the full extrinsic curvature tensor $[K_{ab}]=0$. This automatically ensures $[K_{\tau\tau}]=0$ alongside the angular components, leaving no residual constraint (see Appendix~\ref{app:derivation} for explicit verification). Under these circumstances, the junction dynamics are fully determined by the angular component $K_{\theta\theta}$.

\subsection{Calculation of $K_{\theta\theta}$ and Matching}

For the static interior region $\mathcal{M}^{-}$, using the normalization $u_\mu u^\mu=-c_0^2$, the angular component of the extrinsic curvature is obtained as (see Appendix~\ref{app:derivation})
\begin{equation}
K_{\theta\theta}^{-}
=R\sqrt{1-\frac{2G_0M}{c_0^2R}+\frac{\dot{R}^2}{c_0^2}}\,,
\label{eq:K_in}
\end{equation}
where $R$ denotes the physical radius and $\dot{R}=dR/d\tau$ is the proper velocity of the shell.

For the GCT exterior $\mathcal{M}^{+}$, the unit normal vector to the comoving boundary $\chi=\chi_b$ is purely spatial. Since the spatial metric depends only on the scale factor $a(t)$, the unit normal vector takes the form $n_\mu^{+}=(0,a(t),0,0)$ in the orthonormal frame. Using the Christoffel symbol $\Gamma^{\chi}_{\theta\theta}=-\chi$, the angular component of the extrinsic curvature is
\begin{equation}
K_{\theta\theta}^{+}
=-\Gamma^{\chi}_{\theta\theta}n_{\chi}
= -(-\chi_b)a(t)
=a(t)\chi_b
=R\,.
\label{eq:K_out}
\end{equation}

Imposing the junction condition $[K_{\theta\theta}]=0$ therefore yields
\begin{equation}
R
=R\sqrt{1-\frac{2G_0M}{c_0^2R}+\frac{\dot{R}^2}{c_0^2}}\,,
\end{equation}
which leads, after squaring and simplification, to
\begin{equation}
\dot{R}^2=\frac{2G_0M}{R}\,.
\label{eq:junction_eq}
\end{equation}

\subsection{Relation to the GCT Friedmann Background}

The proper velocity $\dot{R}$ can be related to the cosmological expansion using the time transformation in Eq.~(\ref{eq:time_out}),
\begin{equation}
\dot{R}
=\frac{dR}{dt_{+}}\frac{dt_{+}}{d\tau}
=HR\,a(t_{+})^{-b/4}
=HR\,\frac{c_0}{c(t_{+})}\,,
\end{equation}
where $H=\dot{a}/a$ is the Hubble parameter. Substituting this expression into Eq.~(\ref{eq:junction_eq}) gives
\begin{equation}
\left(HR\frac{c_0}{c(t_{+})}\right)^2
=\frac{2G_0M}{R}\,.
\end{equation}

Introducing the mean density $\rho_m=M/(\frac{4}{3}\pi R^3)$, the relation becomes
\begin{equation}
\frac{H^2}{c(t_{+})^2}
=\frac{8\pi G_0}{3c_0^2}\rho_m\,.
\label{eq:gct_friedmann_si}
\end{equation}
Equation~(\ref{eq:gct_friedmann_si}) should be interpreted as a geometric consistency relation arising from the junction construction. When combined with the dynamical scalings of $c(t)$, $G(t)$, and $\rho_m(a)$ provided by the GCT framework~\cite{Lee:2020zts,Lee:2025osx}, it reproduces the standard GCT-Friedmann relation. In this sense, the Friedmann-like form obtained here is not presented as a new independent dynamical field equation, but as the shell condition rewritten in density variables.

This structure is closely related to the mass-compensation condition familiar from Einstein--Straus matching. In that standard setting one has
\begin{equation}
R_\Sigma(t)=a(t)\,r_b
\qquad\text{and}\qquad
M=\frac{4\pi}{3}\rho\,R_\Sigma^3 \, ,
\end{equation}
so that the spherical vacuole boundary expands with the background even though the interior region is locally static. The present construction shares this geometric feature when the boundary is taken to be comoving. The narrower point established here is different: once the exterior metric carries the time-dependent lapse $N_+(t)=a^{b/4}$, the junction consistency condition acquires the explicit factor $c_0/c(t_+)$, so that the matching probes the compatibility of distinct temporal normalizations rather than merely the continuity of matter variables across the boundary.

\begin{quote}
\textbf{Geometric Consistency Condition (GCC).}

\textit{
A smooth Israel junction between a Schwarzschild interior and a GCT-FLRW exterior is possible only if the background expansion satisfies the GCT-modified Friedmann relation. This condition expresses a necessary compatibility between a locally static region with fixed clocks and a cosmological region with a time-dependent lapse.
}
\end{quote}

It is worth clarifying that the geometric decoupling derived here does not imply any modification of local relativistic physics, including black-hole thermodynamics or laboratory clock rates. Within the interior region $\mathcal{M}^-$, the metric admits a timelike Killing vector normalized by the static lapse $N_-=1$, so that local notions of energy, temperature, and proper time coincide with those of standard general relativity. Consequently, quantities such as the Hawking temperature, horizon entropy, and atomic transition frequencies are determined entirely by the local Schwarzschild geometry and are insensitive to the cosmological evolution of the exterior lapse.

The time-dependent lapse $N_+(t)$ therefore affects only the embedding of the local region into the global spacetime and the relation between local proper time and cosmological observables, rather than the internal dynamics of the virialized system.

This derivation shows that locally constant values of $c_0$ and $G_0$ are compatible with a cosmological background in which the lapse function varies with time. The resulting separation between local physical measurements and the global cosmological time normalization therefore arises as a geometric property of the spacetime construction.

\section{Discussion}
\label{sec:discussion}

The analysis in the preceding section establishes an idealized geometric setting in which a locally static interior can be matched to a time--dependent cosmological background within the GCT framework under the simplifying comoving-boundary ansatz. In particular, the junction calculation shows that the continuity of the extrinsic curvature restricts the class of admissible background evolutions. This result may be interpreted as a statement of geometric consistency: only those cosmological histories that satisfy the GCT--modified Friedmann relation can be matched smoothly to a region with fixed local time normalization. It is also worth noting that related matching problems arise in contexts where the signature of the metric changes, for example in spacetimes containing transitions between Euclidean and Lorentzian regions. Matching conditions for such signature-changing geometries have been discussed in the literature~\cite{Ellis:1991sp,Kerner:1993fm}. 

\subsection{Geometric Matching and Local Consistency}

In many modified--gravity scenarios, the reconciliation of large--scale departures from $\Lambda$CDM with local tests of gravity relies on dynamical screening mechanisms, such as the chameleon or Vainshtein effects \cite{Khoury:2003aq,Vainshtein:1972sx}. These approaches introduce additional degrees of freedom whose dynamics suppress observable deviations in high--density environments.

The construction examined here instead highlights a geometric aspect of spacetime matching: local and cosmological regions may be described by distinct time normalizations that remain mutually consistent through the IJCs.

The present analysis has been carried out in the idealized limit in which the surface stress--energy tensor on the matching hypersurface vanishes, $S_{ab}=0$. In this limit, the junction conditions isolate the purely geometric content of the construction. More general situations in which finite boundary layers are present may lead to additional effects that resemble dynamical screening, but the $S_{ab}=0$ case provides a useful baseline in which the role of geometry can be examined without introducing further microphysical assumptions.

\subsection{Time Normalization and Physical Duality of Clocks}

The geometric construction also bears on the interpretation of dimensionful constants such as $c$ and $G$ \cite{Duff:2014mva}. Within the present framework, the distinction is not between physical and unphysical variations, but between different time normalizations associated with local and cosmological frames.

In the interior region $\mathcal{M}^{-}$, the lapse takes the constant value $N=1$, and locally measured quantities correspond to fixed parameters $c_0$ and $G_0$. In the exterior region $\mathcal{M}^{+}$, the lapse evolves as $N\propto a^{b/4}$, leading to time--dependent background quantities $c(t)$ and $G(t)$.

The IJCs ensure that these two descriptions can be embedded consistently in a single spacetime. In this sense, the apparent variation of dimensionful quantities in cosmological descriptions may be interpreted as a consequence of the time normalization associated with the cosmological frame, rather than as a modification of locally measured physics.

\subsection{Virialized Systems and Local Physics}

The geometric construction studied here is motivated by virialized astrophysical systems such as galaxy clusters and black holes, but the present paper does not claim to provide a fully realistic model of such systems. Under the comoving ansatz,
\begin{equation}
R(\tau)=a\bigl(t_+(\tau)\bigr)\chi_b \, ,
\end{equation}
the matching surface has a time-dependent areal radius, so the construction should be interpreted as an idealized zeroth-order geometric model rather than the exact boundary worldtube of a static bound object. Its role is to isolate how a locally static time gauge can remain consistent with an exterior cosmological region described by a different temporal normalization.

While realistic systems may deviate from perfect spherical symmetry or possess non-comoving and slowly evolving boundaries, the construction presented here corresponds only to the leading-order monopole component of the matching problem. Corrections associated with non-sphericity or boundary dynamics, including the more general case $\chi=\chi(\tau)$, are deferred to future work \cite{Lee:PartII}.

In regions where the metric approaches a static Schwarzschild form, the lapse tends to unity, and the spacetime admits a timelike Killing vector. Local notions of time, energy, and temperature are then those of standard general relativity. Quantities such as atomic transition frequencies, laboratory clock rates, and black--hole thermodynamic relations are determined by the local geometry and are insensitive to the cosmological evolution of the exterior lapse. While the present work focuses on the geometric consistency of
spacetime matching, related issues concerning the averaging of inhomogeneous cosmologies and the fitting problem have also been
widely discussed in the literature~\cite{Ellis:1987zz,Buchert:1999er,Clarkson:2011zq}.

\subsection{Observational Considerations}

The construction developed here suggests that cosmological observables probing the Hubble flow and null geodesics are sensitive to the background lapse $N(t)$, while local measurements performed in virialized environments probe a static time normalization.

Observable quantities such as CTD, luminosity distances, or integrated expansion histories may therefore depend on the background lapse scaling $N(t)\propto a^{b/4}$, even if the local microphysics of astrophysical sources remains unchanged.

More concretely, in the GCT framework the relation between boundary proper time and exterior cosmological time takes the form
\begin{equation}
-c_0^2 d\tau^2=-c(t_+)^2 dt_+^2,
\qquad
\frac{dt_+}{d\tau}=\frac{c_0}{c(t_+)}=a^{-b/4}(t_+),
\label{eq:obs_time_relation}
\end{equation}
so that the propagation of temporal signals through the cosmological background is modified relative to the standard FLRW case when $b\neq 0$. Correspondingly, the cosmological time-dilation scaling is shifted from the standard $(1+z)$ form to
\begin{equation}
\Delta t_{\rm obs}\propto (1+z)^{\,1-b/4}\,\Delta t_{\rm emit},
\label{eq:obs_td_relation}
\end{equation}
which provides a concrete example of an observable quantity affected by the evolving lapse at the phenomenological level.

At the same time, to avoid overstating the scope of the present work, we emphasize that Eqs.~(\ref{eq:obs_time_relation}) and (\ref{eq:obs_td_relation}) are given here only to illustrate the type of invariant observational consequence that can arise in the GCT framework. A detailed confrontation with data lies beyond the scope of the present geometric analysis and is deferred to companion studies.

By contrast, precision experiments carried out in gravitationally bound systems are governed by the local static gauge and are not expected to show secular drifts in fundamental constants.

The empirical implications of this framework arise from comparing cosmological observations that probe the global propagation of light with local measurements performed in gravitationally bound systems.

The geometric distinction between global cosmological time and locally measured proper time also provides a possible framework for interpreting several recent observational results related to CTD. In particular, the GCT framework predicts that transient phenomena originating from compact or strongly bound systems can behave as effective ``geometric clocks'' tracing the propagation of light through the cosmological background, while persistent sources governed by thermal emission processes may exhibit different observational signatures. This perspective has been applied to the interpretation of time-dilation signals in Type Ia supernovae and gamma-ray bursts, the absence of clear time-dilation trends in quasar variability studies, and possible connections between CTD and the inference of the Hubble constant~\cite{Lee:2023ucu,Lee:2024kxa,Lee:2025vha,Lee:2026kyz}. In this sense, the geometric matching construction presented here provides the spacetime foundation for a broader phenomenological program investigating the observational consequences of cosmological time reparameterization.  

The apparent variation of dimensional quantities such as $c(t)$ in the GCT framework reflects the choice of global cosmological time slicing rather than a breakdown of local physical laws. In particular, precision spectroscopic measurements probe atomic systems residing within gravitationally bound environments, whose local metric is effectively decoupled from the cosmological background. As a result, dimensionless quantities such as the fine-structure constant remain invariant in the local proper frame even if the background lapse function evolves. This environmental decoupling mechanism has been discussed in detail in~\cite{Lee:2023xfg}.

\section{Conclusion}
\label{sec:conclusion}

This work has examined the compatibility between a time--dependent cosmological lapse and the existence of locally static gravitational systems within the Generalized Cosmological Time (GCT) framework. Using the Israel junction conditions, a composite spacetime was constructed in which a Schwarzschild interior is matched to a GCT--FLRW exterior with $N(t)=a^{b/4}$ under the simplifying comoving-boundary ansatz.

The requirement of extrinsic curvature continuity leads to a Friedmann--type relation, Eq.~(\ref{eq:gct_friedmann_si}), which coincides with the background equations of motion of the GCT framework. This relation should be interpreted conservatively as a geometric consistency condition: only those cosmological histories that satisfy the GCT--modified Friedmann relation can be matched smoothly to a region with a fixed local time normalization in the present zeroth-order construction. The junction does not determine the cosmological dynamics by itself, nor does it by itself provide a fully realistic model of a virialized boundary.

Within the interior region, the presence of a timelike Killing vector normalized by a unit lapse ensures that local notions of time, energy, and temperature coincide with those of standard general relativity. Local physical processes, including laboratory clocks and black--hole thermodynamics, are governed by the interior geometry and are insensitive to the time normalization of the exterior region. Any apparent variation of dimensionful quantities in the GCT framework may thus be associated with the choice of global time coordinate rather than with a breakdown of local physical laws.

Taken together, these results show that the coexistence of cosmological time--reparameterization effects and locally invariant physics can be analyzed within a single spacetime geometry, without introducing additional propagating degrees of freedom. In this perspective, geometric matching provides a useful zeroth-order framework for studying how distinct cosmological and local time normalizations may coexist. The realistic non-comoving virialized-boundary problem, however, is left for future work.

While the present work establishes the zeroth-order geometric matching conditions, further testing this framework on data surfaces---particularly in the context of null horizons, characteristic evolution, and Cauchy surfaces for dynamical black holes---remains an important direction for future investigation.

\appendix

\section{Detailed Derivation of Junction Conditions in SI Units}
\label{app:derivation}

Throughout this appendix we adopt the coordinate convention
\begin{equation}
x^0 = c_0 t ,
\end{equation}
where $c_0$ denotes the locally measured speed of light.  This choice ensures that the metric components remain dimensionless, while allowing the time--dependent quantity $c(t)$ appearing in the GCT framework to be distinguished from the locally measured constant $c_0$.

The four--velocity normalization is taken as
\begin{equation}
u_\mu u^\mu=-c_0^2 ,
\end{equation}
which corresponds to the standard SI definition of proper time.

The extrinsic curvature of the hypersurface $\Sigma$ is defined as
\begin{equation}
K_{ab} = n_{\mu;\nu} e^\mu_a e^\nu_b ,
\end{equation}
where $n_\mu$ is the unit normal vector to $\Sigma$ and $e^\mu_a = \partial x^\mu/\partial y^a$ are the tangent vectors to the hypersurface coordinates $y^a=(\tau,\theta,\phi)$.

\subsection{Proof of $[K_{\tau\tau}]=0$ for $S_{ab}=0$}

The Israel junction conditions read
\begin{equation}
[K_{ab}] - h_{ab}[K] = - \frac{8 \pi G}{c^4} S_{ab}.
\end{equation}
For a pressureless boundary with no surface energy density, we set
\begin{equation}
S_{ab}=0 .
\end{equation}
Taking the trace gives
\begin{equation}
[K]=0 .
\end{equation}
Substituting back yields
\begin{equation}
[K_{ab}]=0 .
\end{equation}
In particular,
\begin{equation}
[K_{\tau\tau}]=0 .
\end{equation}
Thus, the absence of a surface stress tensor implies that the extrinsic curvature must be continuous in all directions across the hypersurface.

\subsection{Interior: Static Schwarzschild Spacetime}

The interior metric is

\begin{equation}
ds_{-}^2=-f(r)c_0^2 dt_{-}^2+\frac{1}{f(r)}dr^2+r^2 d\Omega^2 ,
\end{equation}
with
\begin{equation}
f(r)=1-\frac{2G_0M}{c_0^2 r}.
\end{equation}
The junction hypersurface is parametrized by
\begin{equation}
r=R(\tau).
\end{equation}
The four--velocity is
\[
u^\mu=(c_0\dot t_-,\dot R,0,0).
\]
The normalization condition
\begin{equation}
g_{\mu\nu}u^\mu u^\nu=-c_0^2
\end{equation}
gives
\begin{equation}
-f(c_0\dot t_-)^2+\frac{1}{f}\dot R^2=-c_0^2 .
\end{equation}
Solving for $\dot t_-$ gives
\begin{equation}
\dot t_-=\frac{1}{f}\sqrt{f+\frac{\dot R^2}{c_0^2}} .
\end{equation}
The unit normal vector $n_\mu$ is determined by the conditions
\begin{equation}
n_\mu u^\mu = 0 , \qquad n_\mu n^\mu = 1 .
\end{equation}
Writing

\[
n_\mu=(-A,B,0,0)
\]
the orthogonality condition yields

\[
A=\dot R , \qquad B=c_0\dot t_- .
\]
Thus
\[
n_\mu=\Big(-\frac{\dot R}{c_0},\,\dot t_-,\,0,\,0\Big).
\]
The normalization is verified using

\begin{equation}
n_\mu n^\mu
=-\frac{\dot R^2}{f c_0^2}+f\dot t_-^2
=1 .
\end{equation}

The angular component of the extrinsic curvature becomes
\begin{equation}
K_{\theta\theta}^-=-\Gamma^r_{\theta\theta}n_r .
\end{equation}
Using
\[
\Gamma^r_{\theta\theta}=-r f(r)
\]
we obtain
\begin{equation}
K_{\theta\theta}^-
=R f\dot t_- .
\end{equation}
Substituting $\dot t_-$ yields
\begin{equation}
K_{\theta\theta}^-
=R\sqrt{1-\frac{2G_0M}{c_0^2R}+\frac{\dot R^2}{c_0^2}} .
\end{equation}

\subsection{Exterior: GCT--FLRW Spacetime}

The exterior metric is
\begin{equation}
ds_{+}^2 =-c(t)^2dt_{+}^2 +a(t)^2 d\chi^2 +a(t)^2\chi^2 d\Omega^2 .
\end{equation}
The hypersurface is defined by
\[
\chi=\chi_b .
\]
The unit normal vector is therefore purely spatial,
\[
n_\mu^+=(0,a(t),0,0).
\]
Using
\[
\Gamma^\chi_{\theta\theta}=-\chi
\]
the extrinsic curvature becomes
\begin{equation}
K_{\theta\theta}^+ =-\Gamma^\chi_{\theta\theta}n_\chi =a(t)\chi_b .
\end{equation}
Using
\[
R=a(t)\chi_b
\]
we obtain

\begin{equation}
K_{\theta\theta}^+=R .
\end{equation}

\subsection{Second Junction Condition}

The condition
\begin{equation}
K_{\theta\theta}^-=K_{\theta\theta}^+
\end{equation}
gives
\begin{equation}
R\sqrt{1-\frac{2G_0M}{c_0^2R}+\frac{\dot R^2}{c_0^2}}=R .
\end{equation}
After simplification
\begin{equation}
\dot R^2=\frac{2G_0M}{R}.
\end{equation}
Using
\[
\frac{dt_+}{d\tau}=\frac{c_0}{c(t)}
\]
one obtains
\[
\dot R = HR \frac{c_0}{c(t)} .
\]
Thus
\begin{equation}
\frac{H^2}{c(t)^2}
=\frac{8\pi G_0}{3c_0^2}\rho_m .
\end{equation}
This relation demonstrates that the Israel junction condition leads to a Friedmann--type relation when expressed in terms of the time--dependent speed of light $c(t)$ appearing in the GCT framework.

\section{Extension to Schwarzschild--de Sitter Interior}
\label{app:sds}

To include a cosmological constant, the interior region may be described by a static Schwarzschild--de~Sitter geometry with
\begin{equation}
f(r)=1-\frac{2G_0M}{c_0^2 r}-\frac{\Lambda r^2}{3}\,.
\end{equation}
Following the same steps as above, the interior extrinsic curvature becomes
\begin{equation}
K_{\theta\theta}^-
=R\sqrt{1-\frac{2G_0M}{c_0^2R}-\frac{\Lambda R^2}{3}+\frac{\dot R^2}{c_0^2}}\,.
\end{equation}
Matching to $K_{\theta\theta}^+=R$ gives
\begin{equation}
1-\frac{2G_0M}{c_0^2R}-\frac{\Lambda R^2}{3}+\frac{\dot R^2}{c_0^2}=1\,.
\end{equation}
With $\dot R=HR\,c_0/c(t)$, this yields
\begin{equation}
\frac{H^2}{c(t)^2}
=\frac{8\pi G_0}{3c_0^2}\rho_m+\frac{\Lambda}{3}\,.
\end{equation}
This demonstrates that the same geometric matching applies when the interior spacetime is generalized to include a cosmological constant.

\section*{Acknowledgments}
This work is supported by the Basic Science Research Program through the National Research Foundation of Korea (NRF), funded by the Ministry of Science and ICT under Grant No.~NRF-RS-2021-NR059413 and NRF-2022R1A2C1005050.

\section*{Data Availability Statement}
No new data were created or analyzed in this study.

\section*{Conflict of Interest}
The author declares no competing interests.


\end{document}